\documentclass[epj,twocolumn]{webofc}
\usepackage[varg]{txfonts}   % Web of Conferences font
\usepackage{graphicx}

\woctitle{Powders \& Grains 2017}

\usepackage{color}

\begin{document}

\title{Granular rheology: measuring boundary forces with laser-cut leaf springs}

\author{\firstname{Zhu} \lastname{Tang}\inst{1}\fnsep\thanks{\email{ztang2@ncsu.edu}} \and
        \firstname{Theodore A.} \lastname{Brzinski}\inst{1,2}\fnsep\thanks{\email{tbrzinski@haverford.edu }} \and
        \firstname{Karen E.} \lastname{Daniels}\inst{1}\fnsep\thanks{\email{kdaniel@ncsu.edu}}
        % etc.
}

\institute{Department of Physics, North Carolina State University, Raleigh, NC 27695 USA; 
\and
           Department of Physics, Haverford College, Haverford, PA 19041 USA
          }

\abstract{In granular physics experiments, it is a persistent challenge to obtain the boundary stress measurements necessary to provide full a rheological characterization of the dynamics. Here, we describe a new technique by which the outer boundary of a 2D Couette cell both confines the granular material and provides spatially- and temporally- resolved stress measurements.  This key advance is enabled by desktop laser-cutting technology, which allows us to design and cut linearly-deformable walls with a specified spring constant. By tracking the position of each segment of the wall, we measure both the normal and tangential stress throughout the experiment. This permits us to calculate the amount of shear stress provided by basal friction, and thereby determine accurate values of  $\mu(I)$. 
%I deleted the details about $mu(I)$ here, to de-emphasize that and ``save'' it for your real paper
%I added what was new about these walls
%I took out the negative statements about photoelastic measurements. The goal of this paper is to report a new method, not to denigrate a previous one (particularly one that you are about to start using.
}

\maketitle 

%\KED{Zhu, please take a look at what I added and subtracted throughout the paper, paragraph by paragraph. Please talk with me about any of them you disagree with or don't understand.}

%\KED{Zhu, for many papers, you will need to submit 1 file for each figure, so please learn to use software for doing this. In this case, we just submit a single PDF so it is okay (don't change anything). That won't be true for the paper you're working on.}

%\KED{Please don't mix .eps and .pdf in the same file: LaTeX has to make extra files in order to work around the mix of those two.}

%\KED{Pressure is always isotropic (it's the symmetric part of the stress tensor), so you don't want to say ``normal pressure'' even though it is the pressure that provides the normal stress on the boundary. Let's discuss if this isn't clear. You also want to use either ``tangential stress'' or ``shear stress'' consistently throughout. }

\section{Introduction}  % =================================================================
 
It is an open question what constitutive equations best describe flows of dense cohesionless granular materials \cite{forterre2008flows,Kamrin2012,Bouzid2013}. There has come to be a consensus that two dimensionless parameters play a key role: interial number $I$ and the friction $\mu$. Each of these can be defined at the particle scale. The inertial number is given by
\begin{equation}
I \equiv \frac{|\dot{\gamma}|d}{\sqrt{P/\rho}}
\label{eq:I}
\end{equation}
where $\rho$ is density of the solid granular material, $d$ is their diameter, $\dot{\gamma}$ is the local shear rate, and $P$ is the local pressure.  Higher values of $I$ correspond to rapid flows, and lower values to slower flows. 
The ratio
\begin{equation}
\mu \equiv \frac{\tau}{P}
\label{eq:mu}
\end{equation}
is the local ratio of tangential stress $\tau$ to the normal stress (pressure $P$), and $\mu(I)$ is observed to be a good empirical descriptor of the rheology of the system \cite{forterre2008flows} for well-developed flows. However, both  $\dot{\gamma}$ and $\mu$ can exhibit strong spatial and temporal gradients. Recently, non-local extensions to the $\mu(I)$ rheology \cite{Kamrin2012,Bouzid2013} aim to provide a theoretical framework for capturing  such features as the  transition from inertial to creeping flow \cite{koval2009annular}, boundary-driven shear-banding \cite{Midi2004,fenistein2003kinematics,Cheng2006}, and fluidization due to non-local perturbations \cite{nichol2010flow,reddy2011evidence,wandersman2014nonlocal}.
%added specific examples of interesting physics: you'll need to finish entering the bibliography entries

\begin{figure}
\centering
\includegraphics[width=0.8\linewidth]{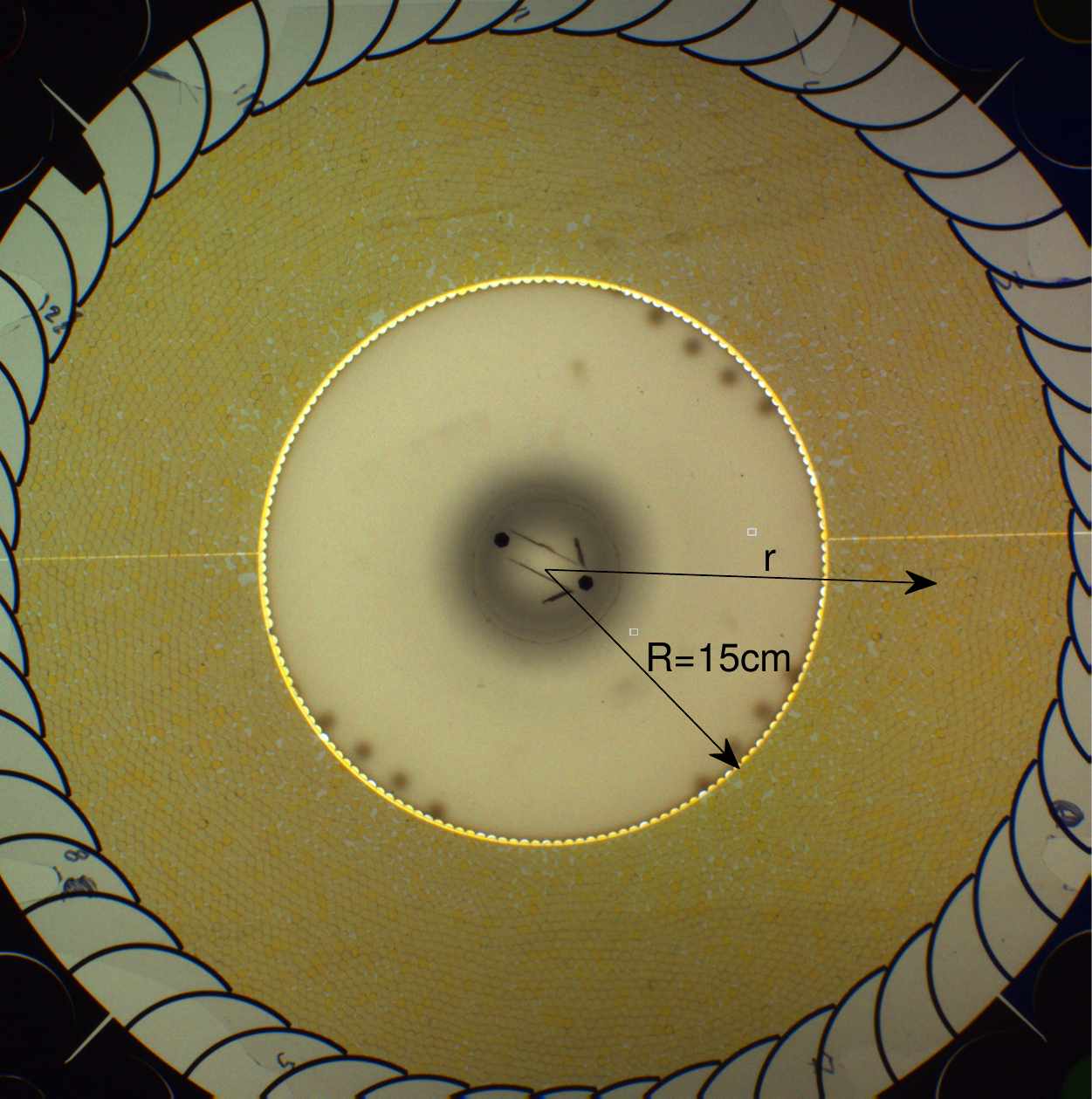}
\caption{Top view of annular Couette experiment with $\approx 5000$ disk-shaped particles confined between a rough inner ring of radius $R_i=15$~cm and a segmented outer ring of radius $R_o=28$~cm. Each of the leaf springs which make up the outer ring can allow dilation under stresses imposed by the dilating granular material. 
\label{fig:exper}}
\end{figure}

However, direct comparisons between experiments and theory have been hampered by the difficulty measuring the tensorial stress within a granular material. Here, we report a new design for a 2D annular Couette cell which can measure the shear and normal stresses at the boundaries (see Fig.~\ref{fig:exper}). This design has several advantages. First, the spring walls can be cut from standard acrylic sheets, making them cheaper and more convenient than using photoelastic particles. Second, the shape of the wall can be easily  customized to have a particular spring constant by changing the thickness or length of the springs. Third, because photoelastic materials are no longer required, experiments on ordinary granular materials are made possible.

Below, we provide a description of the method for making quantitative boundary stress measurements using walls of this type. This involves (1) calibrating a single leaf spring, (2) using image cross-correlation to measure the displacement of each spring tip, and (3)  calculating the stress as a function of time and azimuthal position by combining these two measurements  at multiple positions around the outer wall. We close by presenting sample measurements of the $\mu(I)$ rheology made using this method. 

\section{Method} % =================================================================

\subsection{Apparatus \label{sec:apparatus}}

We develop our technique using a standard annular Couette  geometry, which has the advantage of allowing continuous shear of a granular material to arbitrary total strain. 
The apparatus consists of a rotating inner disk ($R_i=15$~cm) and a fixed outer wall ($R_o=28$~cm) made up of  52 leaf springs which can dilate slightly (a few mm) and thereby provide a stress measurements at the outer wall. The granular material is about 5000 circular and elliptical disks, of diameter $d\approx5$~mm and thickness 3~mm. A photograph of the apparatus from above is shown in Fig.~\ref{fig:exper}.

\subsection{Calibrating the spring wall \label{sec:calibration}}

\begin{figure}
\centering
\includegraphics[width=0.9\linewidth]{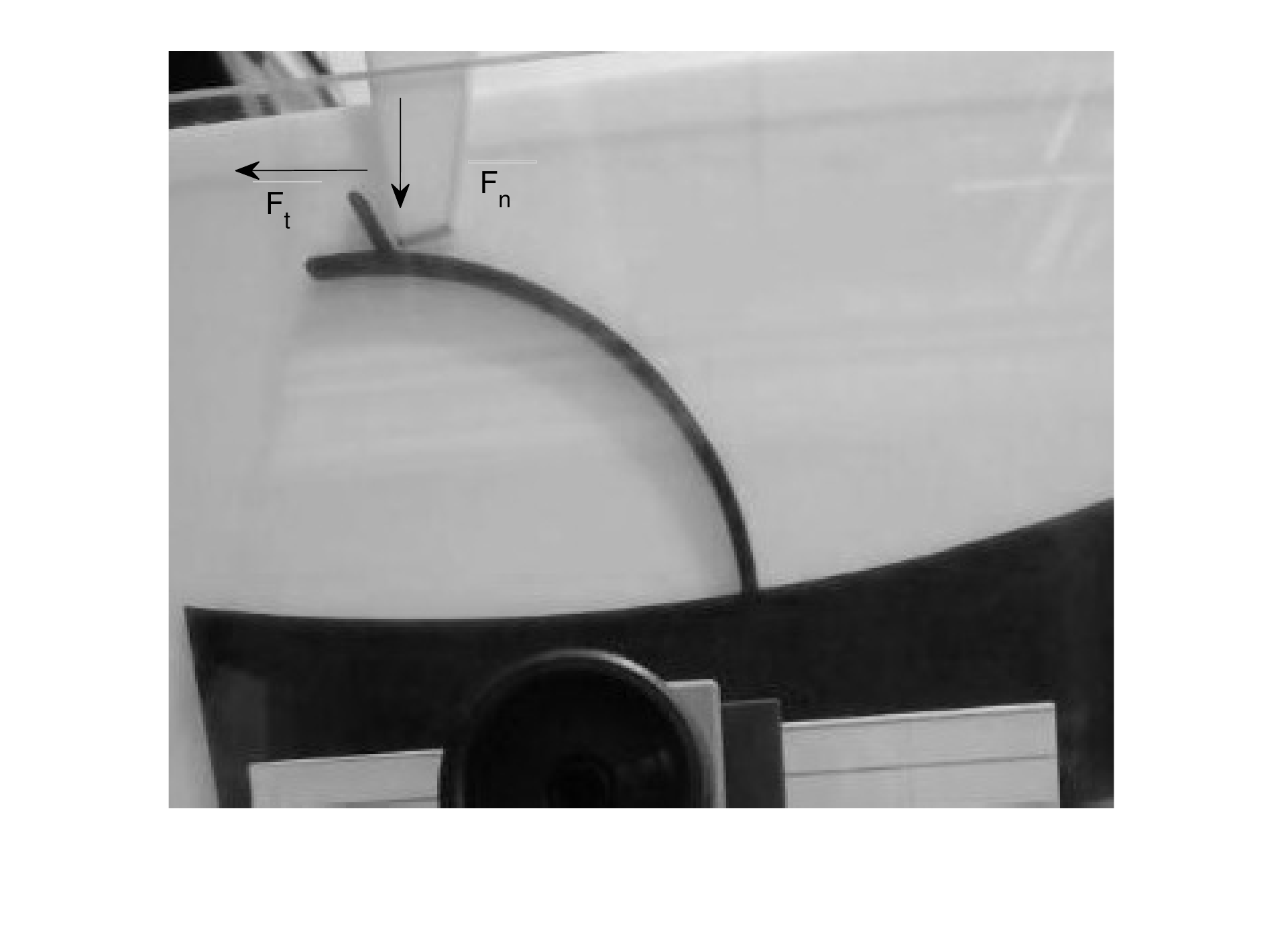}
 \hskip -0.2in (a)
\includegraphics[width=0.9\linewidth]{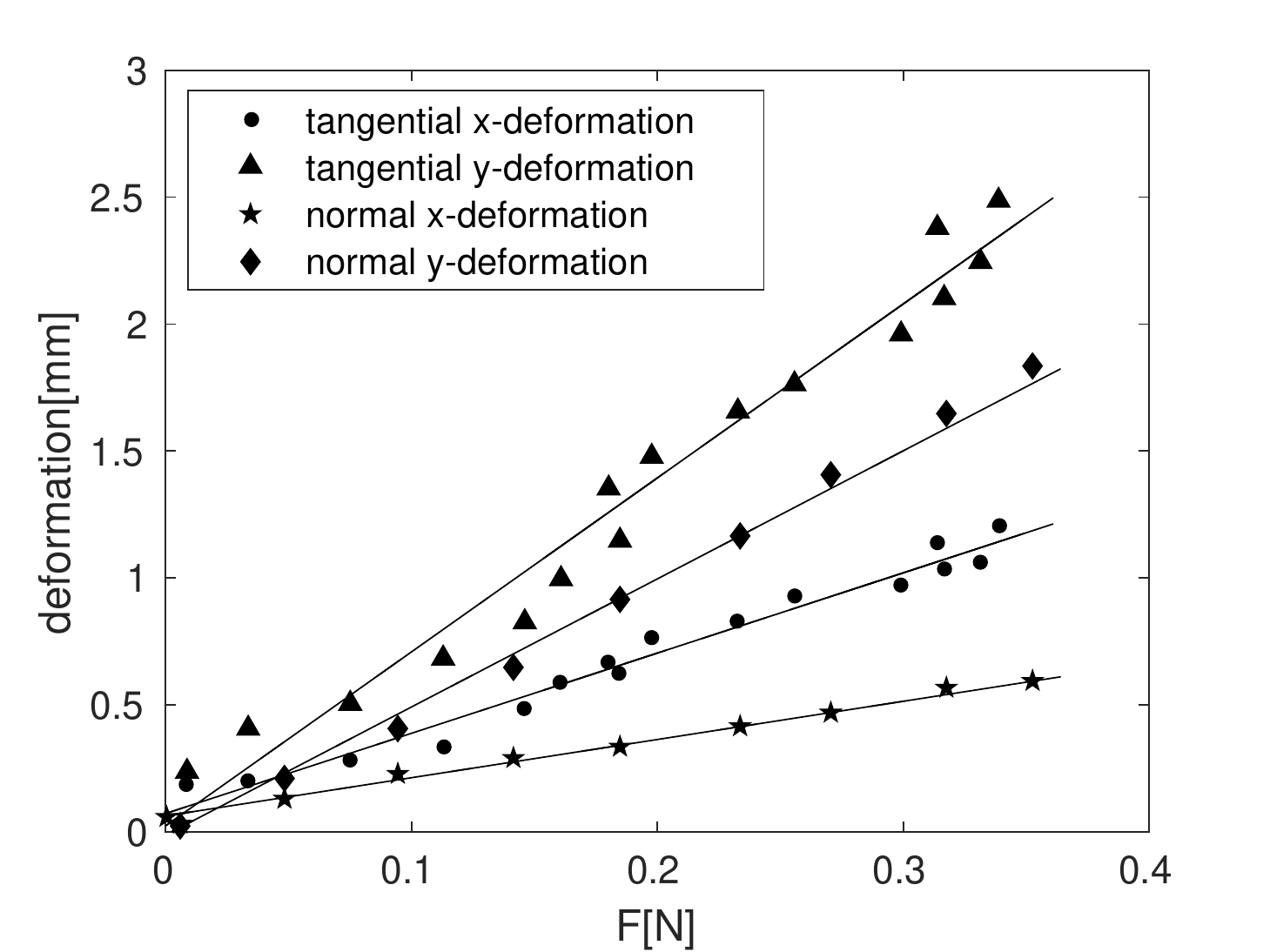}
 \hskip -0.2in (b)
\caption{(a) Photograph of a single leaf spring  mounted in the Instron for calibration, oriented to measure normal forces.
(b) Summary of calibration data for both x- and y-displacements of the tip as a function either pure-normal or pure-tangential applied force. \label{fig:calibration}}
\end{figure}

To provide a calibration for our experiments, we cut a single spring of the same acrylic, but with a short ``handle'' attached, and performed force-displacement measurements using an Instron materials tester. This process is shown in Fig.~\ref{fig:calibration}. One set of measurements was taken with the single spring oriented for normal compression, and a second set with tangential shear, while simultaneously recording a video of the dynamics. In both cases, we measured the x- and y-displacements of the spring tip as a function of applied force and observed a linear response. 
A least-squares fit to the data provides values for the calibration constants. These are 
$C_{n,x}= 1.43 \pm 0.02$~mm/N (x-deformation under normal force), 
$C_{n,y}= 5.06 \pm 0.05$~mm/N (y-deformation under normal force), 
$C_{t,x}= 3.49 \pm 0.05$~mm/N (x-deformation under tangential force), and 
$C_{t,y}= 6.86 \pm 0.08$~mm/N (y-deformation under tangential force). All errors are reported from the fit.

\subsection{Measuring wall deformation  \label{sec:deformation}}

\begin{figure}
\centering
\includegraphics[width=0.78\linewidth,clip]{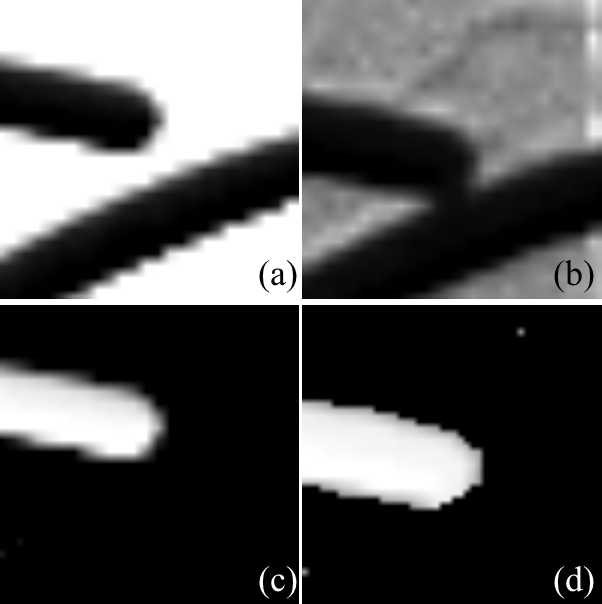}
%\hskip -0.2in \textcolor{white}{(d)}
\includegraphics[width=\linewidth,clip]{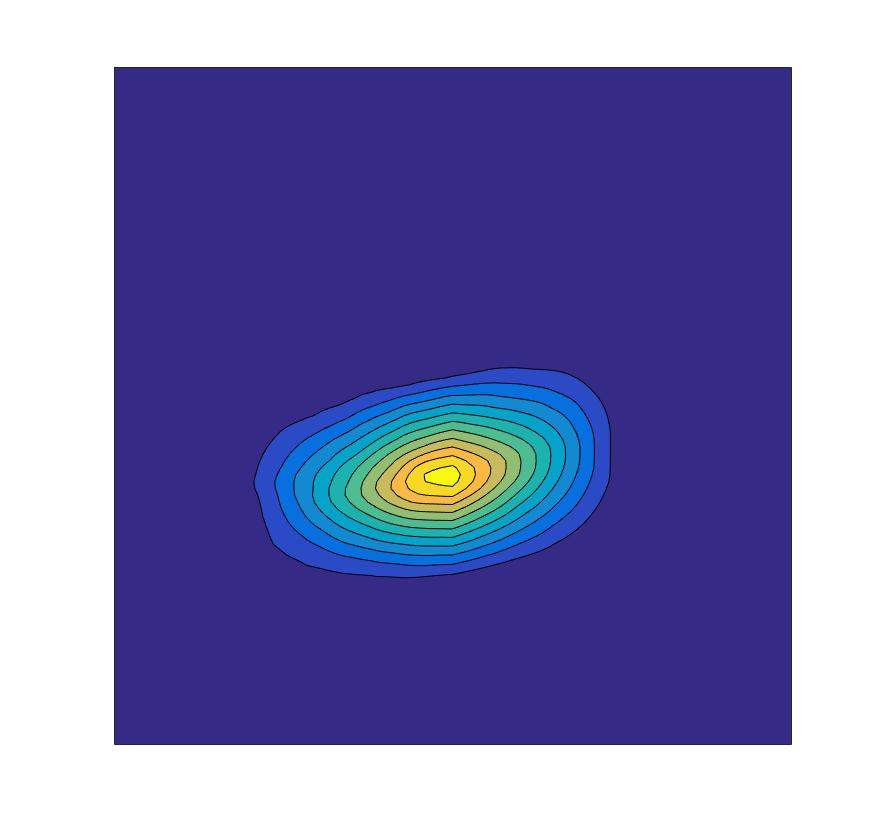}
%\hskip -1in
\vskip -0.55in \textcolor{white}{(e)}
\caption{ Closeup images of the spring tip with (a) and without (b) particles, and these same tips isolated using image-processing (c,d). (e) The contour plot for of cross-correlation values between images (c,d), with the maximum value corresponding to the $(dx,dy)$ displacement.
%The point $\times$ marks the location of the maximum. In this case, this point corresponds to a displacements  $dx=0.41 \pm 0.02$~mm and $dy=2.50 \pm 0.05$~mm, where the error is estimated by taking the standard error over several images.
}

\label{fig:deformation}     
\end{figure}

To determine the normal pressure and the shear stress on the rim, we can make measurements of tip displacements ($dx,dy$, rotated into the appropriate coordinate system) for each of the  leaf springs around the rim. Here, we illustrate the principle using a single leaf spring. First, we extract a subregion of the overhead image (see Fig.~\ref{fig:exper}) in the vicinity of the tip. To determine its displacement, we also extract the same subregion from an image of the experiment taken without particles. Sample images are shown in Fig.~\ref{fig:deformation}ab. However, these raw tip-images also contain a piece of the neighboring spring. Using image-segmentation and masking, we remove the neighboring spring (cd). Using these two images, a simple image cross-correlation (e) determines the  tip displacements $(dx,dy)$ by fitting with subpixel resolution. 

\subsection{Measuring wall stresses \label{sec:wallstress}}

For each measured pair of tip deformations  $(dx,dy)$, we can use the calibration from \S\ref{sec:calibration} to calculate the vector force on the tip: 
\begin{equation}
\begin{array}{lcl}
\displaystyle C_{t,x} F_t + C_{n,x} F_n =dx \\[6pt]
\displaystyle C_{t,y} F_t + C_{n,y} F_n =dy.
\end{array}
\label{eq:convert}
\end{equation}
For the example show in Fig.~\ref{fig:deformation}, this provides $F_n=0.899$~N
%$ \pm 0.002$~N 
and tangential force $F_t=0.251$~N.
%\pm 0.001$~N. The error comes from standard error of $F_t$ and $F_n$ as functions of time.

Since the thickness of the spring wall is $w=3$~mm, and the segment length of each leaf spring is $L=33$~mm, we can convert this to the shear stress $\tau(R_o)$ measured at the outer wall, and pressure $P$:
\begin{equation}
\begin{array}{lcl}
\displaystyle  \tau(R_o)=\frac {F_t}{wL} \\[6pt]
\displaystyle P=\frac {F_n}{wL}.
\end{array}
\label{eq:pressure}
\end{equation}
This provides a segment-averaged estimate for the normal and shear stresses along that particular leaf spring.

\section{Results \label{sec:results}}  % =================================================================

To illustrate how this method can be used to provide rheological measurements, we perform 3 sample runs for two packing fractions ($\phi$) and two rotation rates (specified by the speed $v$ of the inner wall at its rim).
For Case 1 and 2, $\phi =  0.816 \pm 0.003$, and for Case 3, $\phi = 0.840 \pm 0.003$. The error is propaged from errors in the particle and apparatus size measurements.  Case 1 and 3 are taken at $v=0.2$~d/s, and Case 2 at $v=0.02$~d/s, to provide a set of controlled comparisons among the three experiments.

\subsection{Stress measurements} 

\begin{figure}
\raggedright
\includegraphics[width=\linewidth,clip]{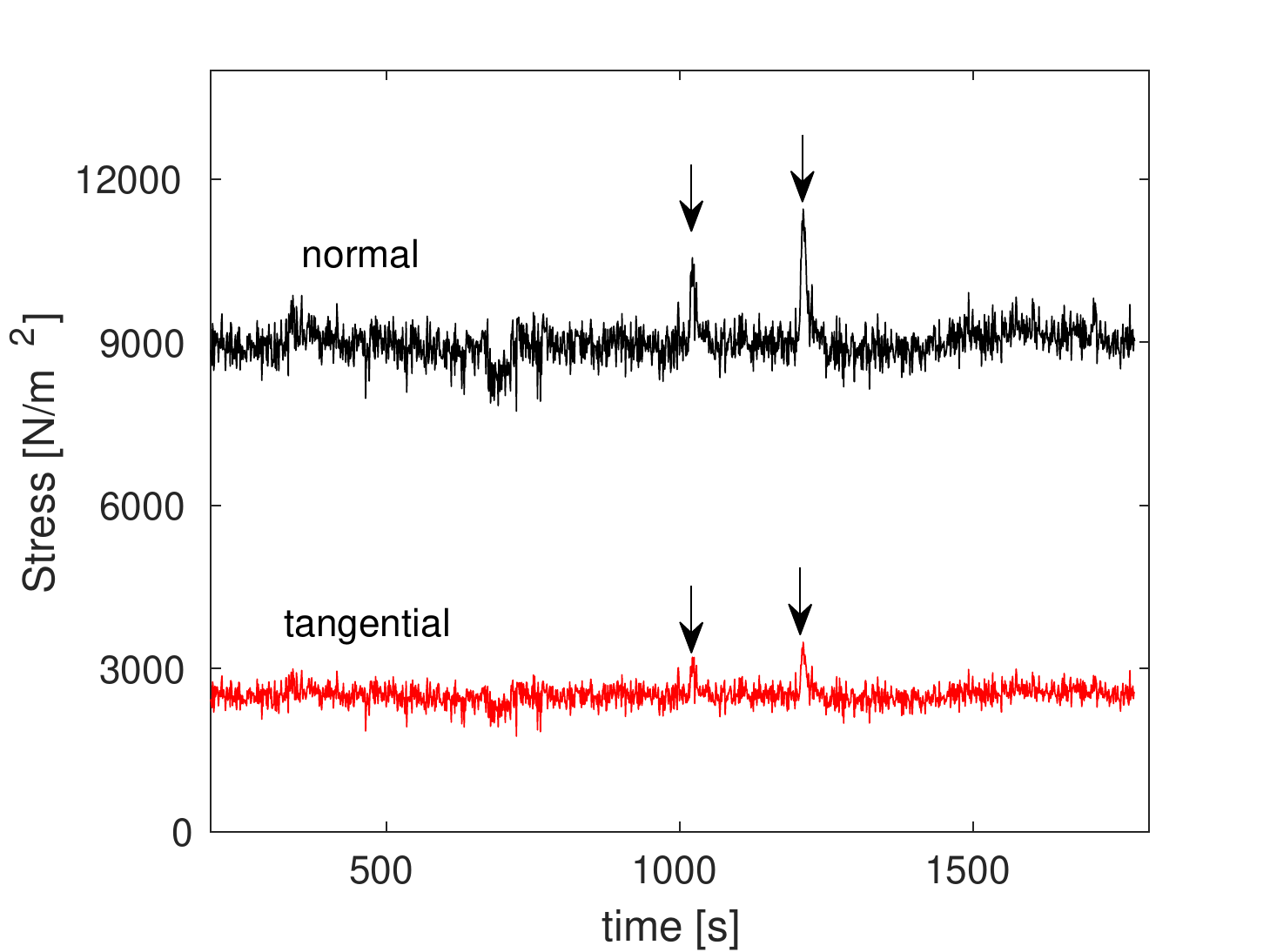}
 \hskip -0.2in (a)
\includegraphics[width=\linewidth,clip]{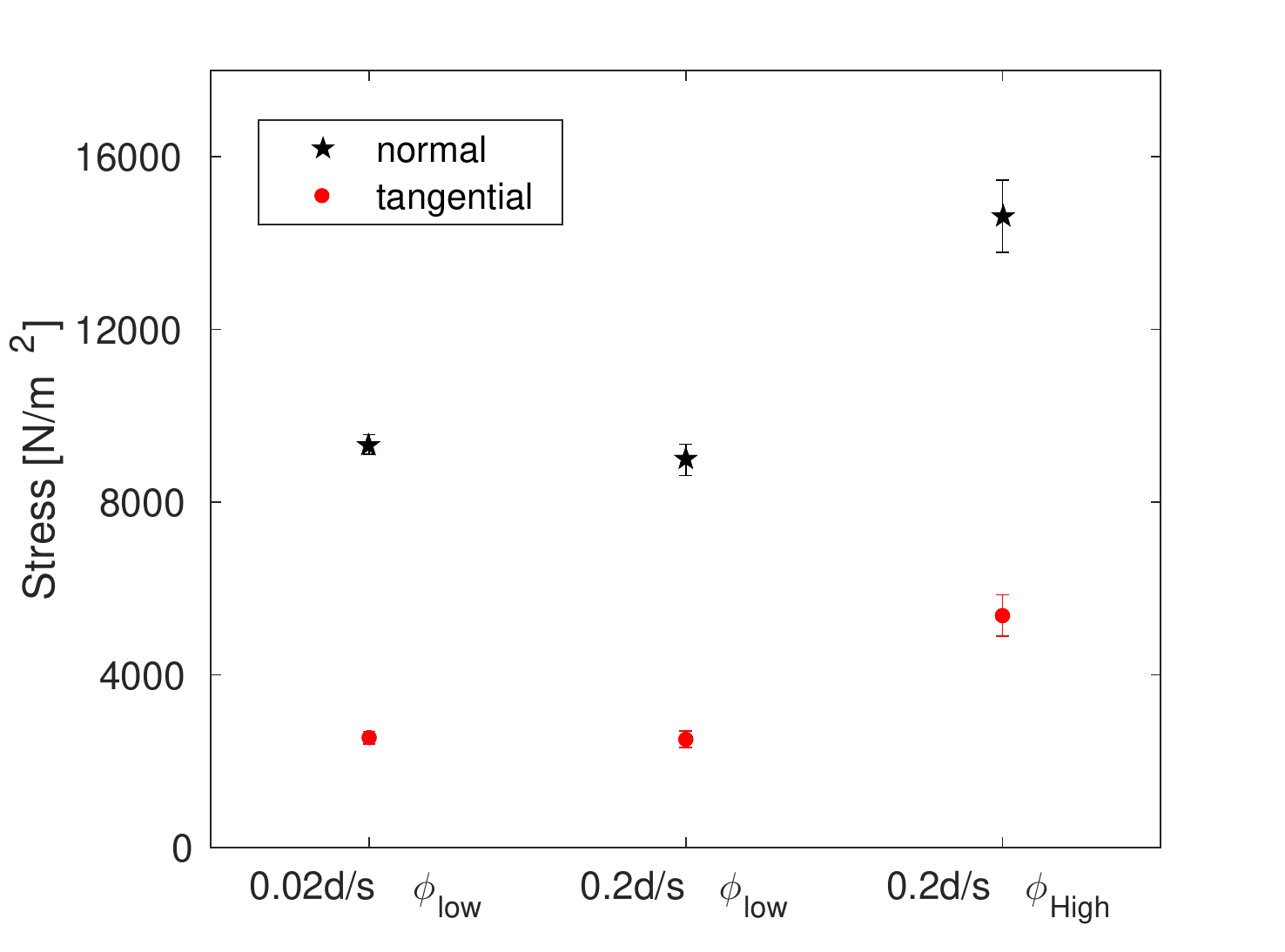}
 \hskip -0.2in (b)
\caption{(a) Time series of normal and tangential stresses for a single leaf spring, Case 1 experiment.
(b) Time-average normal and tangential stress and standard error for all three cases by one sensor.\label{fig:pressure2}
}
\end{figure}

Using the methods of \S\ref{sec:wallstress}, we measure the pressure and the shear stress for a single spring as a function of time (see Fig.~\ref{fig:pressure2}a). We observe that both values fluctuate around a well-defined mean value, punctuated by brief spikes in both. These can be seen near 1000~s and 1200~s, as indicated by the arrows. These are likely due to the transient loading of force chains, which have strong spatial variations on length scales similar to that of a single spring. Measurements at other springs have different average values that this sample, further indicating the presence of spatial heterogeneities. Future work combining boundary measurements with photoelastic measurements are planned. 

For the three sample cases, we can determine how the time-averaged normal and tangential stresses vary according to $\phi$ and $v$. This data is shown in Fig.~\ref{fig:pressure2}b.  We observe that Case 1 and 2 (same $\phi$) have similar values for both normal and tangential stress. We observe that Case 1 and 3 (same $v$), illustrate that both stress values increase with packing fraction, as would be expected.  
\subsection{Rheological measurements \label{sec:muI}}

By combining the wall stress measurements with particle-tracking, we can determine the $\mu(I)$ rheology throughout the granular material. We demonstrate this using values for the two wall stresses measured at  four equally-spaced leaf springs around the outer wall. Improved statistics would be obtained for using all 52 available springs (for which code is under development). In addition, we record the shear stress $S$ measured at the inner wall via a Cooper Instruments torque sensor placed in line with the drive shaft.  

If there were no friction with the supporting plate, the tangential component of the stress would from a maximum at the inner wall according to $\tau(r) =S\left(\frac{R_i}r\right)^2$, where $R_i$ is the radius of the inner wall. Empirically, we  model the effect of this friction with 
\begin{equation}
\begin{array}{lcl}
\displaystyle \tau(r) =S\left(\frac{R_i}{r}\right)^2+\tau_f
\end{array}
\label{eq:tau}
\end{equation}
where  $\tau_f$ is a constant chosen so that $\tau(R_o)$ matches the measured value for the tangential stress at the outer wall. From force-balance, we approximate that the average pressure $P$ is independent of radial position $r$. Thus, the dimensionless stress ratio is $\mu(r)=\tau(r)/P$.

%With the values of the shear stress we obtain from the spring wall, we know the actual value of shear stress at the rim is $\sigma_\mathrm{t}$. The torque sensor on the inner wall can help us calculate the theoretical shear stress at the rim which is $\tau_\mathrm{theory}=S(R/r_\mathrm{rim})^2$, then the shear stress from the basal friction $\tau_0$ can be calculated by $\tau_0=\sigma_\mathrm{t}-\tau_\mathrm{theory}$. So the actual shear stress at different locations 

\begin{figure}
\centering
\includegraphics[width=0.5\textwidth,clip]{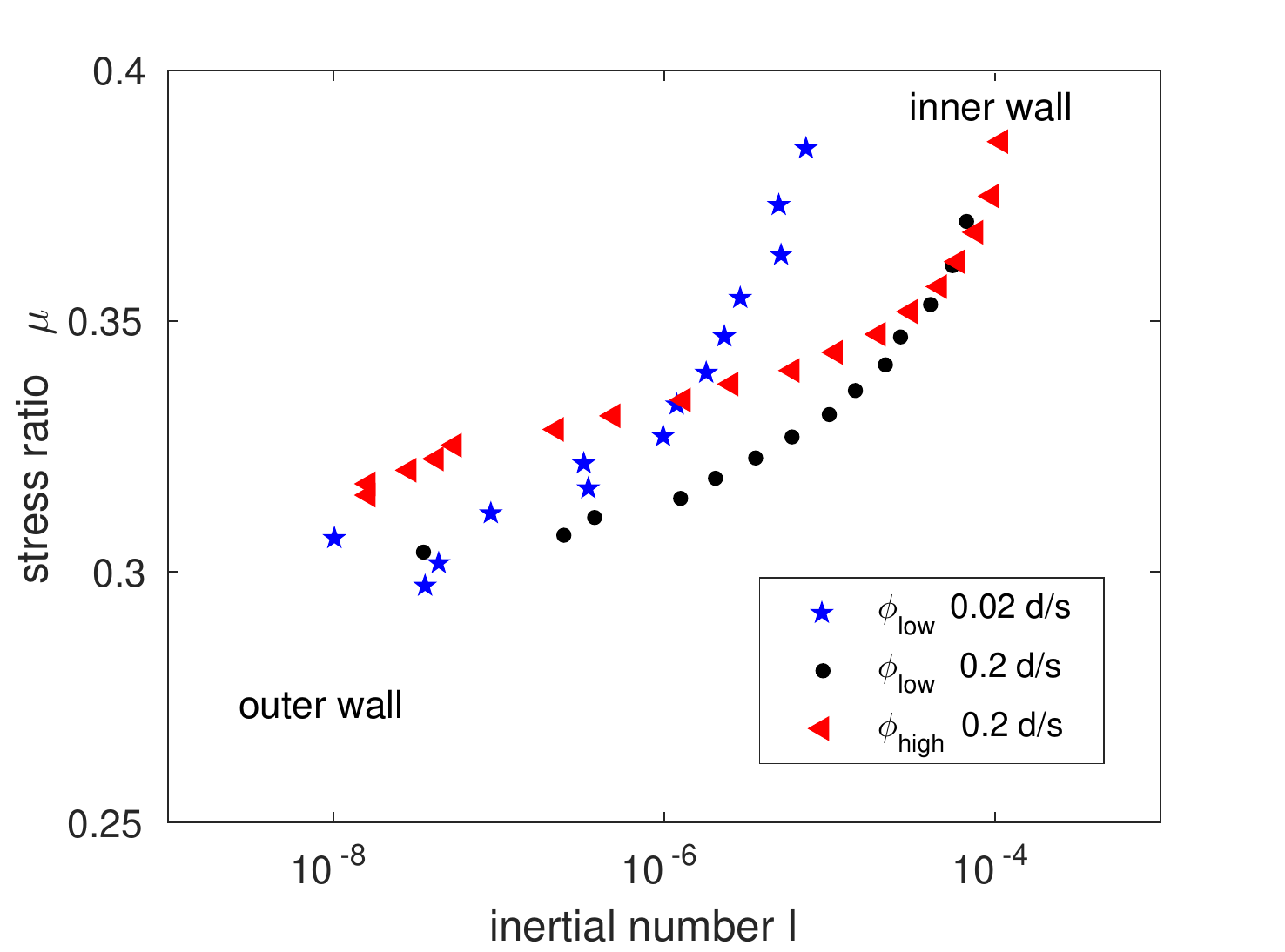}
\caption{Sample $\mu(I)$ curves for all 3 cases.}
\label{fig:rheology}      
\end{figure}

For the three cases, we find the following values for the frictional stress $\tau_f$: Case 1 has  $1870~\mathrm{N/m^2}$; Case 2 has $1630~\mathrm{N/m^2}$, and Case 3 has $3580~\mathrm{N/m^2}$. The errors on all are  $\pm 10$~$\mathrm{N/m^2}$, calculated from the standard error in Fig.~\ref{fig:pressure2}a, averaged over the 4 sensors. 

To measure the inertial number $I$ (Eq.~\ref{eq:I}), we track the particles using video taken at 1 Hz over the whole system  \cite{HT,DB}. We azimuthally-average the velocity profile $v(r)$ and then use Fourier deriviates to calculate 
${\dot{\gamma}}(r) = \frac{1}{2} \left( \frac {\partial v}{\partial r}- \frac{v}{r} \right)$. 
We use the value of $P$ determined from the wall stress measurements, and plot $\mu(r)=\tau(r)/P$ parametrically for all values of $r$.

As shown in Fig.~\ref{fig:rheology}, the $\mu(I)$ rheology depends on both the packing fraction and the rotation rate of the inner disk. We observe that for the same inner wall speed (Case 1 and 3), the $\mu(I)$ curves agree for large $I$ (close to the driving wall). When the inertial number is low, all curves (but particularly Case 1 and 2) approach a constant value $\mu\approx 0.3$ which is dominated by the basal friction.

\section{Conclusion}

We find that laser-cut leaf springs provide a convenient method to both confine a granular material, and measure the boundary wall stresses. In future work, we are expanding the image-processing to provide measurements over the full outer wall, and performing a comparison with photoelastic force measurements. Finally, the $\mu(I)$ measurements provided by Fig.~\ref{fig:rheology}   will allow for quantitative investigation of the utility of nonlocal rheology models \cite{Kamrin2012,Bouzid2013} to describe granular rheology of real materials.

\section*{Acknowledgements}

We thank Michael Shearer, Dave Henann, and Ken Kamrin for useful discussions about the project, and Austin Reid for help creating the boundary wall designs.  We are grateful to the National Science Foundation (NFS DMR-1206808) and International Fine Particle Research Institute (IFPRI) for financial support. 

%\KED{For the bibliography, you do not need to put each reference in its own .bib file: they can all go in the same one. If you are keeping a Mendeley library (or in some other bibliogrpahy software), you can download your .bib file directly rather than entering the information by hand. Please add all of the papers I referenced above. Also, I changed the bibliographystyle to match what they ask for in the template (``woc''), as this is required.}

\bibliographystyle{woc}
\bibliography{rheology}

\begin{thebibliography}{12}

\bibitem{forterre2008flows}
Y.~Forterre, O.~Pouliquen, Annu. Rev. Fluid Mech. \textbf{40}, 1 (2008)

\bibitem{Kamrin2012}
K.~Kamrin, G.~Koval, Physical Review Letters \textbf{108}, 178301 (2012)

\bibitem{Bouzid2013}
M.~Bouzid, M.~Trulsson, P.~Claudin, E.~Cl{\'{e}}ment, B.~Andreotti, Physical
  Review Letters \textbf{111}, 238301 (2013)

\bibitem{koval2009annular}
G.~Koval, J.N. Roux, A.~Corfdir, F.~Chevoir, Physical Review E \textbf{79},
  021306 (2009)

\bibitem{Midi2004}
{GDR Midi}, European Physical Journal E \textbf{14}, 341 (2004)

\bibitem{fenistein2003kinematics}
D.~Fenistein, M.~van Hecke, Nature \textbf{425}, 256 (2003)

\bibitem{Cheng2006}
X.~Cheng, J.B. Lechman, A.~Fernandez-barbero, G.S. Grest, H.M. Jaeger, G.S.
  Karczmar, M.E. Mobius, S.R. Nagel, Physical Review Letters \textbf{96}, 38001
  (2006)

\bibitem{nichol2010flow}
K.~Nichol, A.~Zanin, R.~Bastien, E.~Wandersman, M.~van Hecke, Physical Review
  Letters \textbf{104}, 078302 (2010)

\bibitem{reddy2011evidence}
K.~Reddy, Y.~Forterre, O.~Pouliquen, Physical Review Letters \textbf{106},
  108301 (2011)

\bibitem{wandersman2014nonlocal}
E.~Wandersman, M.~Van~Hecke, EPL (Europhysics Letters) \textbf{105}, 24002
  (2014)

\bibitem{HT}
\emph{Hough transform},
  \url{https://www.mathworks.com/help/images/ref/imfindcircles.html}

\bibitem{DB}
D.~Blair, E.~Dufresne, \emph{The matlab particle tracking code repository},
  \url{http://site.physics.georgetown.edu/matlab/}

\end{thebibliography}

\end{document}